\pdfoutput=1
\documentclass[preprint,preprintnumbers,amsmath,amssymb,floatfix,endfloats*]{revtex4}
\usepackage{graphicx}% Include figure files
\usepackage{dcolumn}% Align table columns on decimal point
\usepackage{bm}% bold math
\usepackage{amsfonts}

\DeclareMathAlphabet{\mathsfsl}{OT1}{cmr}{bx}{it}
\begin{document}
%----------------------------------------------------------------------%
% Title
%----------------------------------------------------------------------%
\title{Accessing a broader range of energy states in metallic glasses by variable-amplitude oscillatory shear}
\author{Nikolai V. Priezjev$^{1,2}$}
\affiliation{$^{1}$Department of Mechanical and Materials
Engineering, Wright State University, Dayton, OH 45435}
\affiliation{$^{2}$National Research University Higher School of
Economics, Moscow 101000, Russia}
\date{\today}
\begin{abstract}

The influence of variable-amplitude loading on the potential energy
and mechanical properties of amorphous materials is investigated
using molecular dynamics simulations. We study a binary mixture that
is either rapidly or slowly cooled across the glass transition
temperature and then subjected to a sequence of shear cycles with
strain amplitudes above and below the yielding strain. It was found
that well annealed glasses can be rejuvenated by small-amplitude
loading if the strain amplitude is occasionally increased above the
critical value. By contrast, poorly annealed glasses are relocated
to progressively lower energy states when subyield cycles are
alternated with large-amplitude cycles that facilitate exploration
of the potential energy landscape. The analysis of nonaffine
displacements revealed that in both cases, the typical size of
plastic rearrangements varies depending on the strain amplitude and
number of cycles, but remains smaller than the system size, thus
preserving structural integrity of amorphous samples.

\vskip 0.5in


\end{abstract}

\maketitle

\section{Introduction}

The fundamental understanding of the interrelationship between the
amorphous structure, mechanical and physical properties of bulk
metallic glasses is important for numerous biomedical and structural
applications~\cite{Wang2012,Trexler10,ZhengBio16}. In contrast to
crystalline materials, where irreversible deformation is controlled
by topological line defects, it was realized that glasses deform
plastically via a sequence of collective rearrangements of groups of
atoms, often referred to as shear transformations~\cite{Spaepen77,
Argon79}. The advantageous properties of metallic glasses include
relatively high strength, large elastic strain limit, high
resistance to corrosion, and biocompatibility, among others, but
they typically suffer from catastrophic failure upon external
deformation, \textit{i.e.}, they are brittle when in a well annealed
state~\cite{Wang2012}. To remediate the latter issue, a number of
processing methods are employed in order to rejuvenate metallic
glasses; for example, cold rolling, high pressure torsion,
irradiation, elastostatic loading, and surface treatments like shot
peening~\cite{Greer16}. More recently, it was found that a
particularly elegant and minimally invasive approach to enhance
potential energy is to thermally cycle glasses between the room and
cryogenic temperatures~\cite{Ketov15,Guo19,Priez18tcyc,
Priez19T2000,Priez19T5000,Mirdamadi19,Priez19one,Jittisa20,
Meylan20,Du20,Guan20}. Despite considerable efforts, however, the
development of efficient processing methods to access a broader
range of energy states in metallic glasses and, at the same time,
maintain their structural integrity remains a challenging problem.

\vskip 0.05in

In the last decades, atomistic simulations have played an important
role in understanding relaxation, rejuvenation and yielding
phenomena in disordered materials subjected to periodic
deformation~\cite{Lacks04,Lo10,Priezjev13,Sastry13,Reichhardt13,
IdoNature15,Shi15,Yang16,Priezjev16,Kawasaki16,Priezjev16a,
Sastry17,Priezjev17,OHern17,Priezjev18,Priezjev18a,Heuer18,
NVP18strload,Sastry19band,PriezSHALT19,ShuoLi19,Peng19, Priez20ba,
Jana20,KawBer20,NVP20altY,Kawasaki20,Priez20del,Priez20heal}. Most
notably, it was discovered that after a certain number of cycles at
zero temperature, disordered solids become locked into the so-called
`limit cycles' where the trajectory of each atom is exactly
reversible during one or more
periods~\cite{Reichhardt13,IdoNature15}. At finite temperatures, the
structural relaxation, sometimes termed as \textit{mechanical
annealing}, proceeds via collective irreversible rearrangements
during hundreds of loading cycles, and the potential energy
gradually approaches a constant
value~\cite{Priezjev18,Priezjev18a,NVP18strload, PriezSHALT19}.
Depending on the initial energy state, it typically takes a number
of transient cycles to yield when the strain amplitude is above the
critical value, and the number of cycles is reduced upon
periodically alternating the loading
direction~\cite{Priezjev17,Priezjev18a,Sastry19band,Priez20ba,
KawBer20,NVP20altY,Priez20del}.  What remains unknown, however, is
how to apply cyclic loading and rejuvenate glasses without the
formation of shear bands, and, on the other hand, how to access
low-energy states in poorly annealed glasses by mechanical
agitation.

\vskip 0.05in

In this paper, the effect of variable-amplitude oscillatory shear
deformation of binary glasses on their energy states and mechanical
properties is investigated using molecular dynamics (MD)
simulations. We consider a model glass former initially cooled well
below the glass transition temperature and then periodically
deformed with strain amplitudes alternating between values below and
above the yielding amplitude.   It will be shown that \textit{well
annealed} glasses become rejuvenated by cyclic loading when the
strain amplitude is once in a while increased above the critical
amplitude, followed by a sequence of low-amplitude cycles.
Remarkably, the same deformation protocol drives \textit{poorly
annealed} glasses to lower energy states as it allows for a more
efficient exploration of the potential energy landscape.

\vskip 0.05in

The rest of this paper is organized as follows. The molecular
dynamics simulations, parameter values, and the deformation protocol
are described in the next section. The results for the potential
energy series under variable-amplitude cycle loading, mechanical
properties, and the spatiotemporal analysis of irreversible
displacements are presented in section\,\ref{sec:Results}.   The
brief summary is given in the last section.

\section{Molecular dynamics simulations}
\label{sec:MD_Model}

In our study, we use the binary (80:20) Lennard-Jones (LJ) mixture
model to represent an amorphous alloy in three dimensions. This
popular model of a glass former was first introduced by Kob and
Andersen (KA), who investigated its structural and dynamical
properties near the glass transition temperature~\cite{KobAnd95}.
The mixture consists of two types of atoms with strongly
non-additive interaction between different types, which suppresses
crystallization upon cooling across the glass transition.
Specifically, any two atoms of types $\alpha,\beta=A,B$ interact via
the truncated LJ potential, as follows:
\begin{equation}
V_{\alpha\beta}(r)=4\,\varepsilon_{\alpha\beta}\,\Big[\Big(\frac{\sigma_{\alpha\beta}}{r}\Big)^{12}\!-
\Big(\frac{\sigma_{\alpha\beta}}{r}\Big)^{6}\,\Big],
\label{Eq:LJ_KA}
\end{equation}
with the parameters: $\varepsilon_{AA}=1.0$, $\varepsilon_{AB}=1.5$,
$\varepsilon_{BB}=0.5$, $\sigma_{AA}=1.0$, $\sigma_{AB}=0.8$,
$\sigma_{BB}=0.88$, and $m_{A}=m_{B}$~\cite{KobAnd95}. A similar
parametrization of the pairwise interaction was used by Weber and
Stillinger to study the amorphous metal-metalloid alloy
$\text{Ni}_{80}\text{P}_{20}$~\cite{Weber85}. In our simulations,
the cutoff radius of the LJ potential was set to
$r_{c,\,\alpha\beta}=2.5\,\sigma_{\alpha\beta}$. The system consists
of $N=60\,000$ atoms. As usual, the simulation results are reported
in terms of the reduced units of length, mass, and energy
$\sigma=\sigma_{AA}$, $m=m_{A}$, and $\varepsilon=\varepsilon_{AA}$.
The equations of motion were solved numerically using the velocity
Verlet algorithm with the time step $\triangle t_{MD}=0.005\,\tau$,
where $\tau=\sigma\sqrt{m/\varepsilon}$ is the LJ
time~\cite{Allen87,Lammps}.

\vskip 0.05in

% equilibration and temperature protocol

The sample preparation protocol involved a thorough equilibration of
the liquid phase at a constant density
$\rho=\rho_A+\rho_B=1.2\,\sigma^{-3}$ in a periodic box of linear
size $L=36.84\,\sigma$.  The system temperature was regulated via
the Nos\'{e}-Hoover thermostat~\cite{Allen87,Lammps}. For reference,
the computer glass transition temperature at $\rho=1.2\,\sigma^{-3}$
is $T_c=0.435\,\varepsilon/k_B$, where $k_B$ is the Boltzmann
constant~\cite{KobAnd95}.  Following the equilibration period, the
binary mixture was cooled to the low temperature
$T_{LJ}=0.01\,\varepsilon/k_B$ with a slow
($10^{-5}\varepsilon/k_{B}\tau$) and fast
($10^{-2}\varepsilon/k_{B}\tau$) cooling rates in order to obtain
well (low energy) and poorly (high energy) annealed samples at
$\rho=1.2\,\sigma^{-3}$.

\vskip 0.05in

% periodic deformation protocol

\begin{figure}[t]
\includegraphics[width=12.0cm,angle=0]{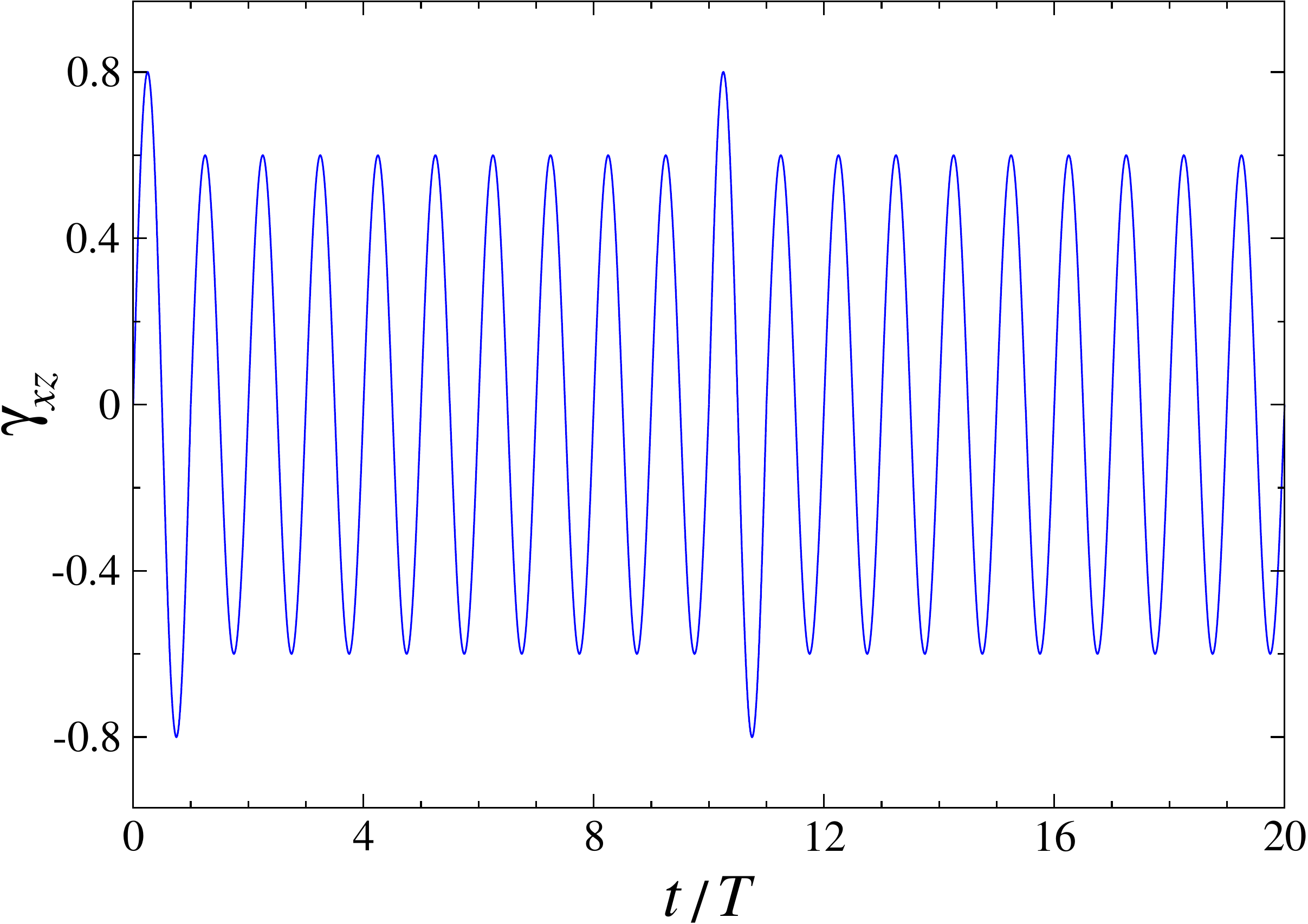}
\caption{An example of the imposed shear strain along the $xz$ plane
during 20 cycles with the oscillation period $T=5000\,\tau$. The
strain amplitude is $\gamma_0=0.06$, except that during every $n$-th
cycle the amplitude is changed to $\gamma_0=0.08$. In this study,
the following values were considered $n=2$, $5$, $10$, $20$, $50$,
and $100$. }
\label{fig:def_var_amp}
\end{figure}

After cooling to $T_{LJ}=0.01\,\varepsilon/k_B$, both samples were
subjected to periodic shear deformation along the $xz$ plane at
constant volume, as follows:
\begin{equation}
\gamma_{xz}(t)=\gamma_0\,\text{sin}(2\pi t/T),
\label{Eq:shear}
\end{equation}
where the oscillation period is $T=5000\,\tau$, and,
correspondingly, oscillation frequency is
$\omega=2\pi/T=1.26\times10^{-3}\,\tau^{-1}$. Unless otherwise
noted, the strain amplitude during $n-1$ cycles is set to
$\gamma_0=0.06$, while the amplitude during every $n$-th cycle is
changed to $\gamma_0=0.08$.  An example of the time dependent shear
strain for $n=10$ is shown in Fig.\,\ref{fig:def_var_amp}. In the
present study, the following values of the periodicity $n$ were
considered $n=2$, $5$, $10$, $20$, $50$, and $100$. The typical
simulation run during 3600 cycles takes about 95 days using 40
processors. During production runs, the potential energy, atomic
positions, and shear strain were stored for post-processing. The
simulations were performed only for one \textit{well annealed} and
one \textit{poorly annealed} samples due to computational
constraints.

\section{Results}
\label{sec:Results}

% intro and brief review

It has been long realized that the atomic structure and mechanical
properties of amorphous alloys strongly depend on the
thermo-mechanical processing history~\cite{Greer16}. In particular,
upon rapid cooling from the liquid state, glasses freeze into highly
unrelaxed configurations and, when externally deformed, exhibit a
gradual crossover from zero to a plateau level of
stress~\cite{Wang2012}. On the other hand, more slowly cooled
glasses settle at lower energy states and become more brittle.
Furthermore, under periodic strain with an amplitude slightly above
a critical value, slowly cooled glasses gradually form a narrow
shear band during a number of transient cycles and then undergo a
yielding transition, whereas rapidly quenched glasses first relax
during hundreds of cycles and then suddenly
yield~\cite{Priezjev17,Sastry17,Sastry19band,Priezjev18a,Priez20ba,NVP20altY}.
By contrast, repeated cycling with an amplitude below the critical
value typically leads to an exploration of progressively lower
energy states via structural
relaxation~\cite{Sastry13,Sastry17,Priezjev18,NVP18strload,PriezSHALT19,Jana20}.
These conclusions were obtained for binary glasses subjected to
cyclic loading at a fixed strain amplitude.

\vskip 0.05in

% intro continued

It can be hypothesized that periodic deformation with the strain
amplitude that is occasionally alternated between values slightly
above and below the critical amplitude, might lead to either
enhanced rejuvenation or accelerated relaxation, depending on the
initial energy state. The rationale for this hypothesis is as
follows.  In the case of \textit{well annealed} glasses, one cycle
with an amplitude slightly larger than the critical value will
increase the potential energy via cooperative irreversible
rearrangements of atoms. Subsequently, one or several subyield
cycles will relax the glass, thus avoiding the formation of
system-spanning shear bands and material failure. Upon iteration,
such deformation protocol might result in steady increase of the
potential energy, and, possibly, a well-defined energy level after
many cycles. On the contrary, \textit{poorly annealed} glasses under
periodic loading below the yielding point tend to relax and
ultimately get trapped in a local minimum of the potential energy
landscape. Thus, one cycle with a large strain amplitude might
relocate the system to an adjacent minimum with lower energy,
leading to accelerated relaxation upon iteration of the sequence of
alternating cycles.

\vskip 0.05in

% intro continued, KA parameters

In our simulations, these scenarios were tested for the well-studied
KA binary mixture model at the low temperature
$T_{LJ}=0.01\,\varepsilon/k_B$ and fixed density
$\rho=1.2\,\sigma^{-3}$.  Thus, it was recently found that the
critical value of the strain amplitude for rapidly cooled KA glasses
at these $T_{LJ}$ and $\rho$ is
$\gamma_0\approx0.067$~\cite{NVP20altY}. Furthermore, well annealed
glasses under periodic shear with the strain amplitude
$\gamma_0=0.08$ were shown to undergo a yielding transition after
about 20 transient cycles, while it might take many cycles to yield
at $\gamma_0=0.07$~\cite{Priezjev17}. Therefore, two values of the
strain amplitude slightly below and above the critical point were
considered; namely, $\gamma_0=0.06$ and $\gamma_0=0.08$. The
deformation protocol consists of one cycle with the strain amplitude
$\gamma_0=0.08$, followed by $n-1$ consecutive cycles with
$\gamma_0=0.06$.  An example of the shear strain variation for the
sequence of alternating cycles, $n=10$, is displayed in
Fig.\,\ref{fig:def_var_amp}. In this study, the simulations were
carried out for the values of the integer $n=2$, $5$, $10$, $20$,
$50$, and $100$. For reference, the results for periodic deformation
with fixed strain amplitudes $\gamma_0=0.06$ and $0.08$ are also
reported.

\vskip 0.05in

% potential energy during cyclic loading; well annealed glass

\begin{figure}[t]
\includegraphics[width=12.0cm,angle=0]{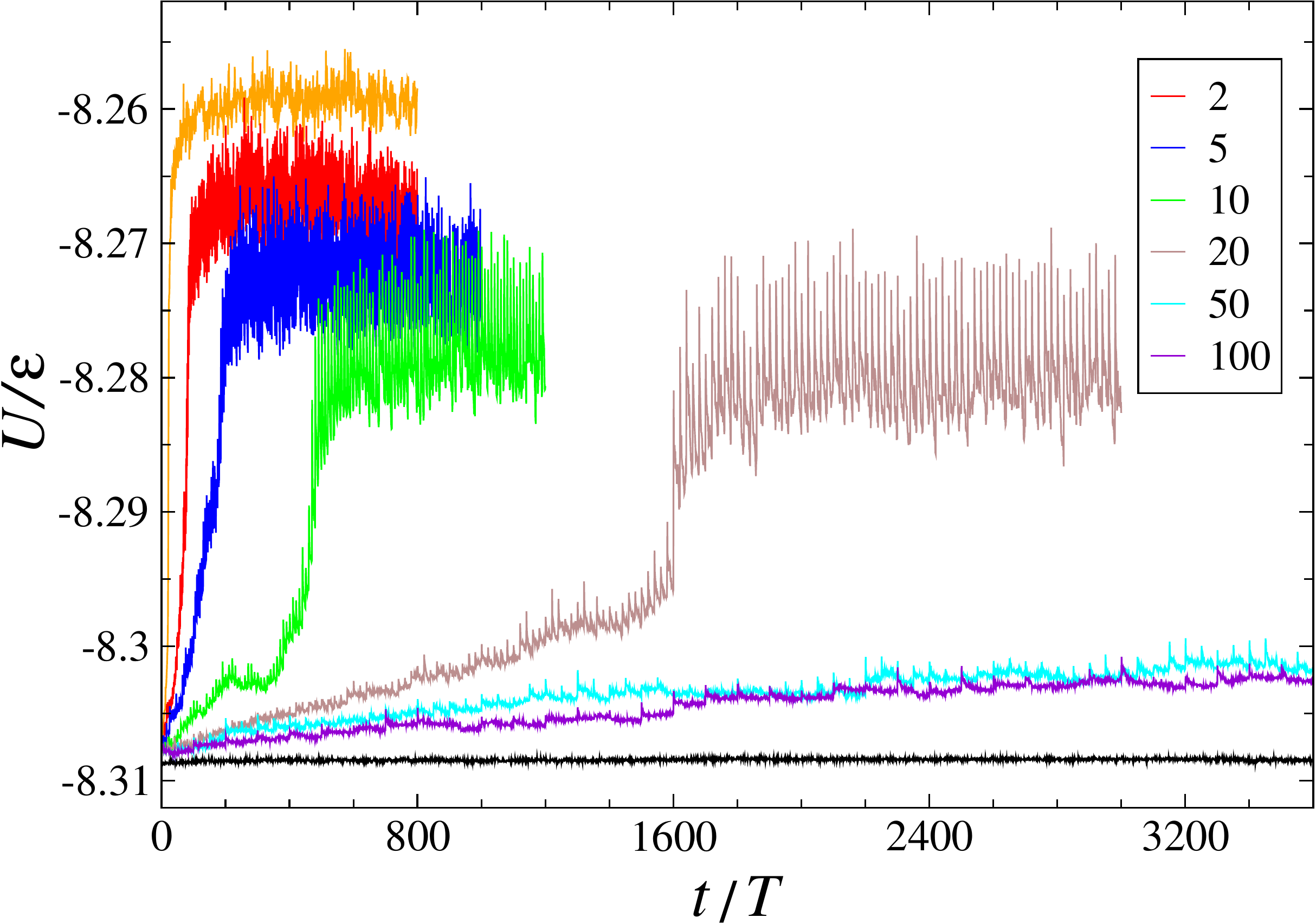}
\caption{(Color online) The potential energy minima at zero strain
as a function of the number of shear cycles with the strain
amplitude $\gamma_0=0.06$ during $n-1$ periods. The strain amplitude
is changed to $\gamma_0=0.08$ during every $n$-th cycle. The values
of periodicity $n$ are listed in the legend. The oscillation period
is $T=5000\,\tau$. The black and orange curves indicate cyclic
loading with the strain amplitudes $\gamma_0=0.06$ and $0.08$,
respectively. The \textit{well annealed} glass was initially
prepared via cooling from the liquid state to
$T_{LJ}=0.01\,\varepsilon/k_B$ with the rate
$10^{-5}\varepsilon/k_{B}\tau$.  }
\label{fig:poten_rejuven}
\end{figure}

The potential energy minima after each cycle at zero strain are
plotted in Fig.\,\ref{fig:poten_rejuven} for the \textit{well
annealed} glass subjected to periodic deformation with periodicity
$n=2$, $5$, $10$, $20$, $50$, and $100$. For comparison, the two
limiting cases with the fixed strain amplitudes $\gamma_0=0.06$ and
$\gamma_0=0.08$ are also included. Note that the potential energy at
$\gamma_0=0.06$ remains essentially constant during 3600 cycles,
indicating that mechanical annealing becomes inefficient for well
annealed glasses even in the presence of thermal fluctuations. It
was previously found that glasses under small-amplitude cyclic shear
cannot access energy states below a certain
threshold~\cite{KawBer20,Priez20del}. Moreover, it can be clearly
observed in Fig.\,\ref{fig:poten_rejuven} that upon increasing
periodicity $n$, the yielding transition is delayed (up to about
1600 cycles for $n=20$). After a sharp increase in potential energy
due to the formation of a shear band across the simulation domain,
and two-phase system continues steady state deformation.

\vskip 0.05in

% potential energy during cyclic loading; well annealed glass; cont

By contrast, the energy curves for the \textit{well annealed} glass
appear to increase and level out for cyclic loading with large
periodicity, $n=50$ and $100$, as shown in
Fig.\,\ref{fig:poten_rejuven}. In these cases, the energy series
resemble a step-like pattern, where a sudden increase after a cycle
at $\gamma_0=0.08$ follows by relaxation during $n-1$ cycles at the
smaller amplitude $\gamma_0=0.06$, thus avoiding the formation of a
shear band.  Hence, these results demonstrate that periodic
deformation with alternating amplitudes can lead to rejuvenation in
the absence of strain localization, provided that periodicity $n$ is
sufficiently large. It can be seen in Fig.\,\ref{fig:poten_rejuven}
that the increase in potential energy after 3600 cycles is about
$0.007\,\varepsilon$, which is comparable to the energy change
reported for the KA binary glass subjected to prolonged elastostatic
loading~\cite{PriezELAST19,PriezELAST20}.

\vskip 0.05in

% potential energy during cyclic loading; poorly annealed glass

\begin{figure}[t]
\includegraphics[width=12.0cm,angle=0]{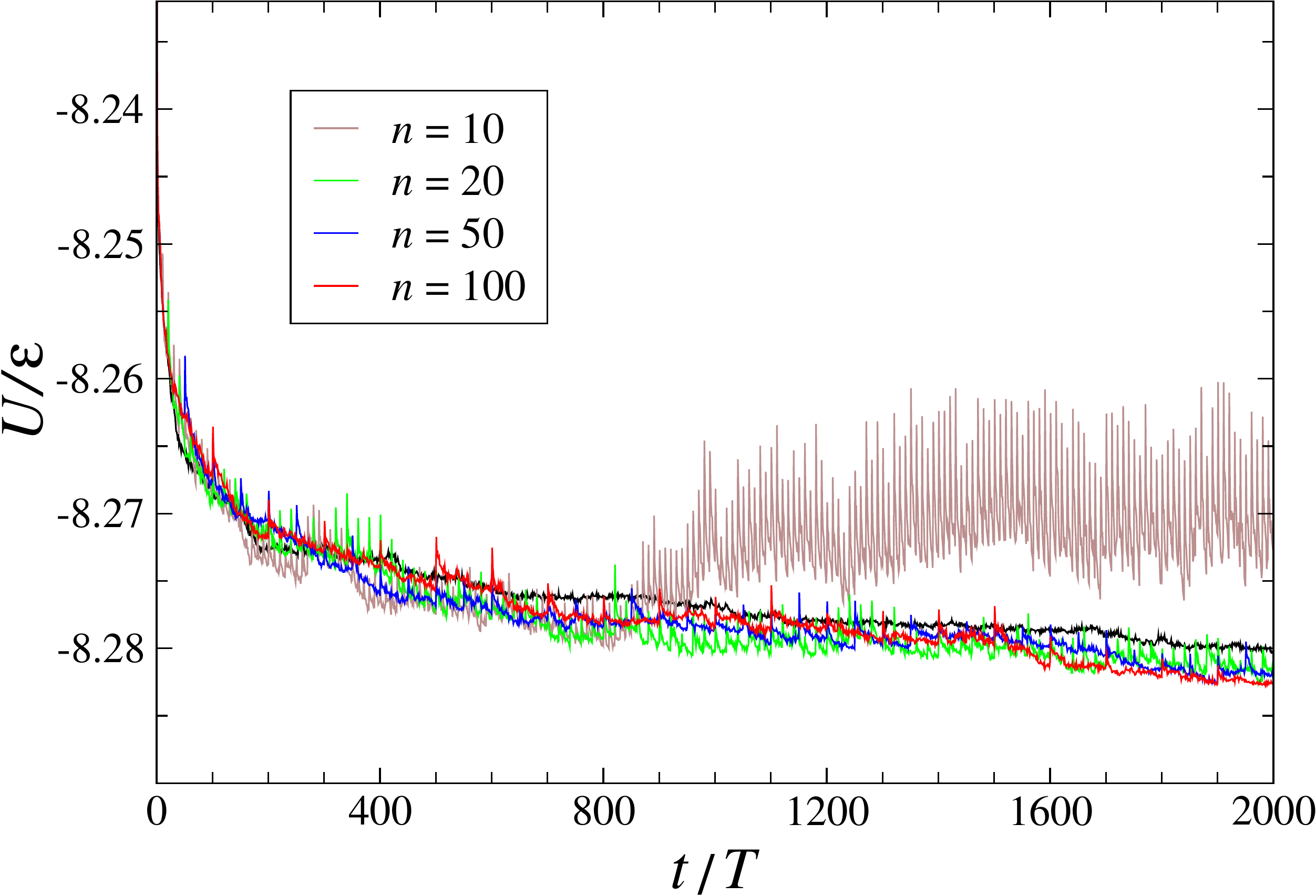}
\caption{(Color online) The time dependence of the potential energy
(at zero strain) for the indicated values of periodicity $n$. The
black curve denotes cyclic loading with the strain amplitude
$\gamma_0=0.06$. The time is expressed in oscillation periods
$T=5000\,\tau$. The \textit{poorly annealed} sample was initially
prepared by cooling with the rate $10^{-2}\varepsilon/k_{B}\tau$ to
$T_{LJ}=0.01\,\varepsilon/k_B$ at $\rho=1.2\,\sigma^{-3}$. }
\label{fig:poten_relax}
\end{figure}

In the case of the \textit{poorly annealed} glass, the potential
energy at the end of each cycle is displayed in
Fig.\,\ref{fig:poten_relax} for the modulated deformation $n=20$,
$50$, $100$ and the fixed strain amplitude $\gamma_0=0.06$.  As is
evident, all deformation protocols initially result in a rapid decay
of the potential energy since many unstable clusters of atoms can be
mechanically driven to lower energy configurations.  When the strain
amplitude $\gamma_0=0.08$ is applied every 10-th cycle, or more
frequently (not shown), the binary glass undergoes a yielding
transition via the formation of a shear band across the system. As
shown in Fig.\,\ref{fig:poten_relax}, the yielding transition for
$n=10$ occurs after about 800 shear cycles when the glass is
mechanically annealed to $U\approx-8.28\,\varepsilon$. These results
are consistent with the critical behavior reported in the previous
MD studies~\cite{Priezjev18a,Sastry19band,Priez20ba,NVP20altY}.
Furthermore, the potential energy continue to gradually decay for
$n=20$, 50, 100, and when the strain amplitude is always fixed at
$\gamma_0=0.06$ (shown by the black curve in
Fig.\,\ref{fig:poten_relax}). It can be seen that upon including a
cycle with the strain amplitude $\gamma_0=0.08$ ($n=20$, 50, and
100), the energy levels are reduced on average by roughly
$0.002\,\varepsilon$ with respect to the black curve. This trend can
be explained by noticing relatively large spikes along the energy
series due to large-scale plastic deformation at $\gamma_0=0.08$,
which increase the probability of relocating the system between
local minima of the potential energy landscape.

\vskip 0.05in

% rejuvenation effect on G and Y; n=100

\begin{figure}[t]
\includegraphics[width=12.0cm,angle=0]{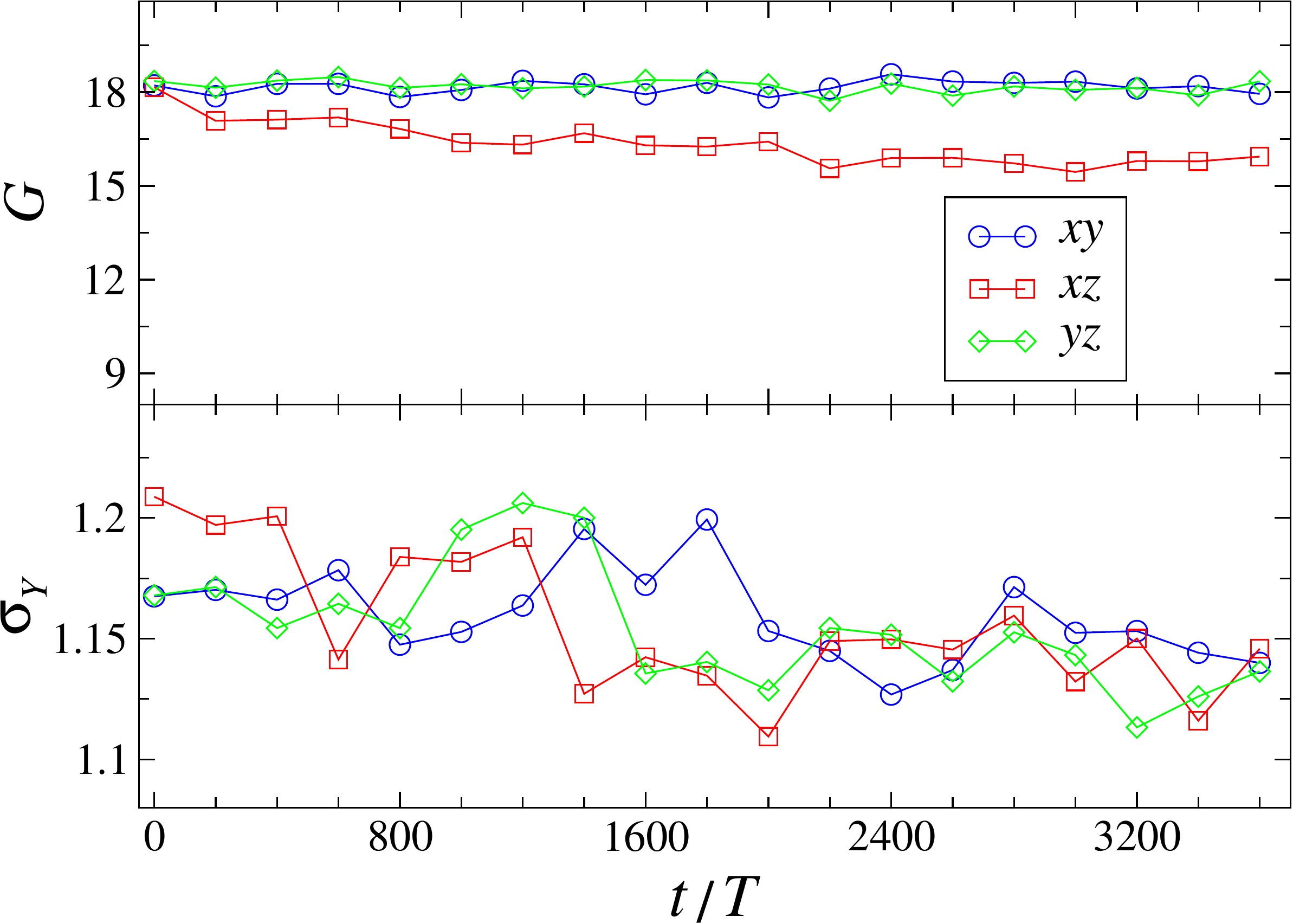}
\caption{(Color online)  The shear modulus $G$ (in units of
$\varepsilon\sigma^{-3}$) and yielding peak $\sigma_Y$ (in units of
$\varepsilon\sigma^{-3}$) versus cycle number for periodic
deformation with $n=100$. The startup continuous shear deformation
with the strain rate $\dot{\gamma}=10^{-5}\,\tau^{-1}$ was imposed
along the $xy$ plane (blue circles), $xz$ plane (red squares), and
$yz$ plane (green diamonds). The variation of the potential energy
for this loading protocol is indicated by the violet curve in
Fig.\,\ref{fig:poten_rejuven}. The glass was initially cooled with
the rate $10^{-5}\varepsilon/k_{B}\tau$ to
$T_{LJ}=0.01\,\varepsilon/k_B$. }
\label{fig:G_and_Y_n100}
\end{figure}

The changes in mechanical properties for the \textit{well annealed}
glass subjected to variable-amplitude deformation with $n=100$ (the
violet curve in Fig.\,\ref{fig:poten_rejuven}) were evaluated by
carrying out mechanical tests at selected time intervals. Thus,
following a certain number of loading cycles, the binary glass,
initially at zero strain, was continuously strained at a fixed rate
$\dot{\gamma}=10^{-5}\,\tau^{-1}$ along the $xy$, $xz$, and $yz$
planes. In each case, the shear modulus $G$ and the peak value of
the stress overshoot $\sigma_Y$ were computed from the linear slope
and the maximum of the stress-strain curve.  The results for $G$ and
$\sigma_Y$ are shown in Fig.\,\ref{fig:G_and_Y_n100} during 3600
loading cycles.  The data are somewhat noisy, since it were
collected for only one realization of disorder, but, nevertheless,
one can clearly see that the shear modulus reduces and acquires
directional anisotropy, and the yielding peak tends to decrease with
the cycle number, which is consistent with gradual rejuvenation
reported in Fig.\,\ref{fig:poten_rejuven}. We also comment that it
was previously shown for cyclically annealed binary glasses that
anisotropy in mechanical properties is reduced when the loading
direction is alternated in two or three spatial
dimensions~\cite{PriezSHALT19}.

\vskip 0.05in

% definition of D2min

We next perform a microscopic analysis of plastic deformation
quantified via the so-called nonaffine displacements of atoms. In
disordered solids, the decomposition of the total displacement of an
atom into affine and nonaffine components can be used to estimate
its relative displacement with respect to neighboring atoms. More
specifically, the nonaffine measure for an atom $i$ can be computed
using the matrix $\mathbf{J}_i$ that linearly transforms a group of
atoms and minimizes the following expression:
\begin{equation}
D^2(t, \Delta t)=\frac{1}{N_i}\sum_{j=1}^{N_i}\Big\{
\mathbf{r}_{j}(t+\Delta t)-\mathbf{r}_{i}(t+\Delta t)-\mathbf{J}_i
\big[ \mathbf{r}_{j}(t) - \mathbf{r}_{i}(t)  \big] \Big\}^2,
\label{Eq:D2min}
\end{equation}
where $\Delta t$ is the time between two configurations, and the
summation is carried over neighbors located within $1.5\,\sigma$
from the position of the $i$-th atom at $\mathbf{r}_{i}(t)$. This
definition was first used by Falk and Langer in the analysis of
localized shear transformations in sheared disordered
solids~\cite{Falk98}. Typically, if the nonaffine displacement of an
atom becomes greater than the cage size, then the local
rearrangement is irreversible, and it is often involves a group of
atoms. In recent years, the spatial and temporal correlations of
nonaffine displacements were extensively studied in amorphous
materials under startup
continuous~\cite{HorbachJR16,Schall07,Pastewka19,Priez20tfic,
Priez19star,Ozawa20,ShiBai20} and time
periodic~\cite{Priezjev16,Priezjev18,Priezjev18a,
PriezSHALT19,Priez20ba,Jana20,NVP20altY,Priez20del} deformation. In
particular, it was demonstrated that the yielding transition is
accompanied by the formation of a system-spanning shear band that
can be clearly identified by plotting atoms with large nonaffine
displacements.

\vskip 0.05in

% nonaffine displacements, well annealed

\begin{figure}[t]
\includegraphics[width=12.0cm,angle=0]{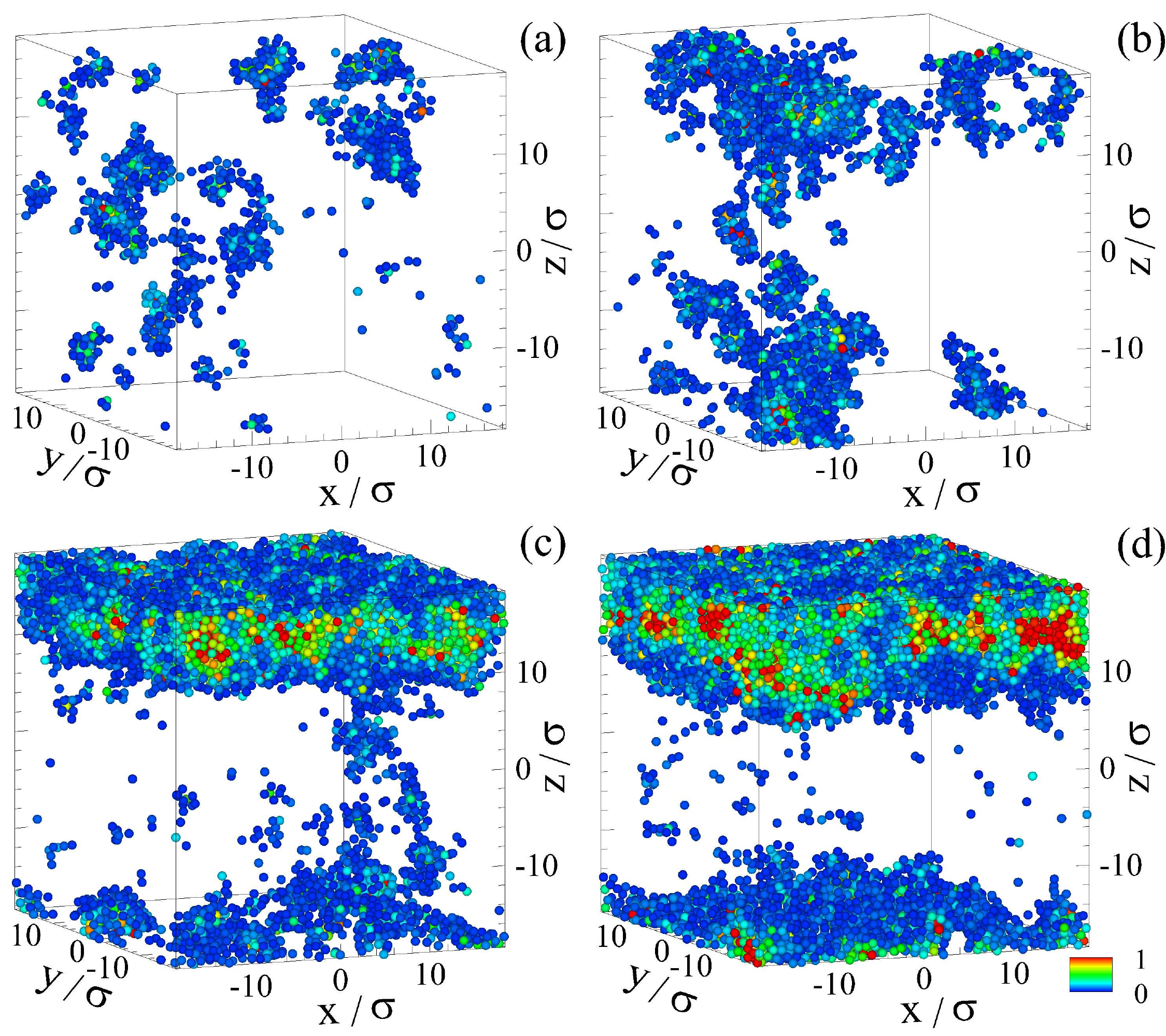}
\caption{(Color online) Snapshots of the \textit{well annealed}
glass subjected to oscillatory shear deformation with periodicity
$n=20$. The potential energy for the same protocol is denoted by the
brown curve in Fig.\,\ref{fig:poten_rejuven}.  The nonaffine
measure, as defined by Eq.\,(\ref{Eq:D2min}), is (a) $D^2(500\,T,
T)>0.04\,\sigma^2$, (b) $D^2(1500\,T, T)>0.04\,\sigma^2$, (c)
$D^2(1600\,T, T)>0.04\,\sigma^2$, and (d) $D^2(2500\,T,
T)>0.04\,\sigma^2$. The oscillation period is $T=5000\,\tau$. The
color in the legend indicates the magnitude of $D^2$.  }
\label{fig:snapshots_rem5_n20}
\end{figure}

\begin{figure}[t]
\includegraphics[width=12.0cm,angle=0]{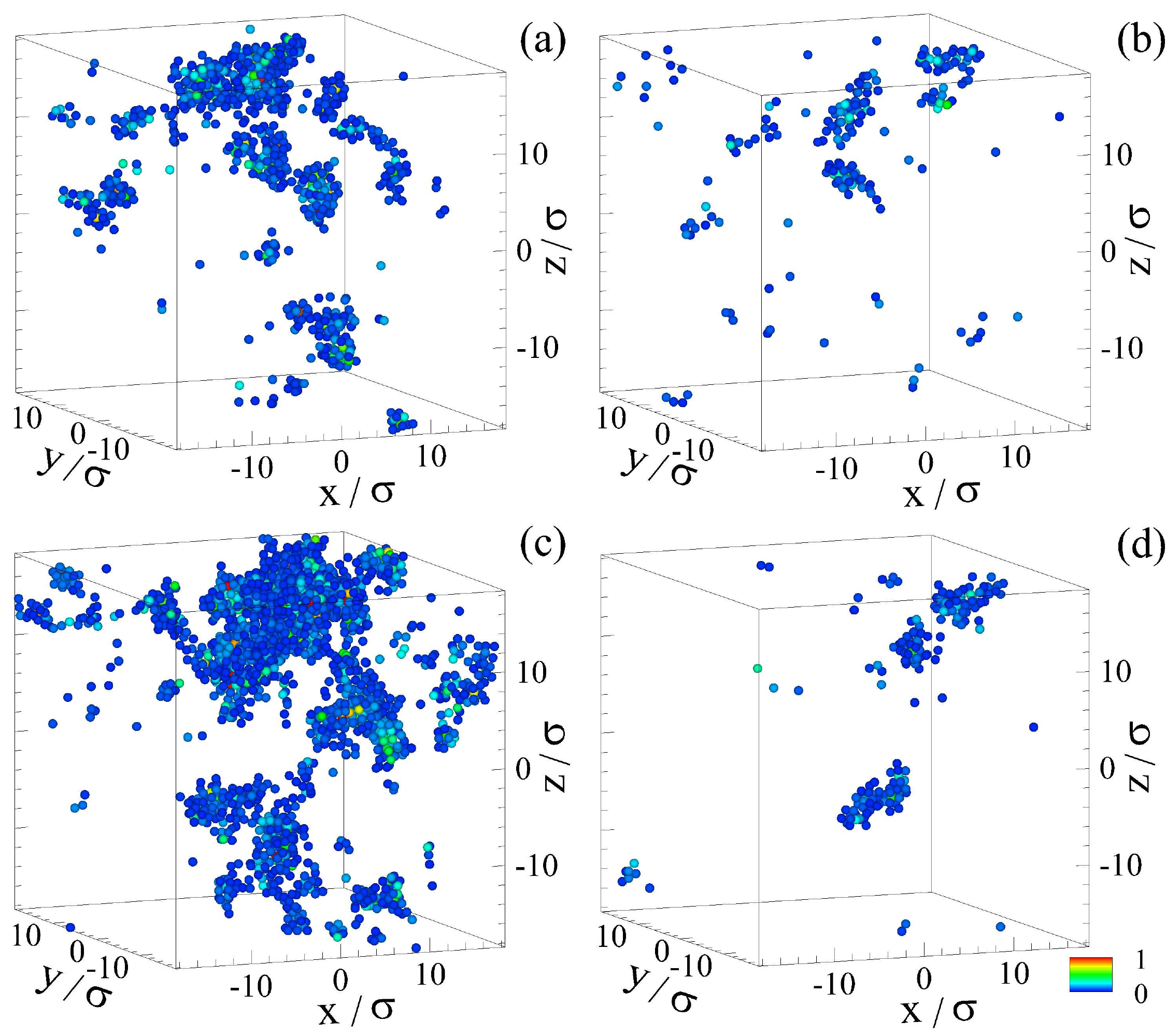}
\caption{(Color online) Plastic rearrangements in the \textit{well
annealed} glass cyclically driven with periodicity $n=100$. The
corresponding energy series are indicated by the violet curve in
Fig.\,\ref{fig:poten_rejuven}. The nonaffine displacements are (a)
$D^2(1000\,T, T)>0.04\,\sigma^2$, (b) $D^2(1050\,T,
T)>0.04\,\sigma^2$, (c) $D^2(3000\,T, T)>0.04\,\sigma^2$, and (d)
$D^2(3050\,T, T)>0.04\,\sigma^2$, where $T=5000\,\tau$. The color
shows the magnitude of $D^2$.  }
\label{fig:snapshots_rem5_n100}
\end{figure}

The atomic configurations of the \textit{well annealed} glass
subjected to variable-amplitude oscillatory deformation are shown in
Fig.\,\ref{fig:snapshots_rem5_n20} for periodicity $n=20$ and in
Fig.\,\ref{fig:snapshots_rem5_n100} for $n=100$. The system
snapshots are taken at selected cycle numbers, and, for clarity,
only atoms with relatively large nonaffine displacements during one
full cycle are displayed.  First, it can be observed in
Fig.\,\ref{fig:snapshots_rem5_n20}\,(a,\,b) that, upon continued
loading, the typical size of clusters of atoms with large nonaffine
displacements increases but remains smaller than the system size.
The appearance of compact clusters during $500$ and $1500$-th
cycles, when the strain amplitude is $\gamma_0=0.08$, correlates
well with the gradual increase in potential energy shown by the
brown curve in Fig.\,\ref{fig:poten_rejuven}.  By contrast, the
sharp increase in potential energy (after about 1600 cycles) signals
a yielding transition and formation of a shear band along the $xy$
plane, which can be clearly seen in
Fig.\,\ref{fig:snapshots_rem5_n20}\,(c,\,d). Second, as illustrated
in Fig.\,\ref{fig:snapshots_rem5_n100}, the situation is
qualitatively different when the strain amplitude is changed to
$\gamma_0=0.08$ less frequently (\textit{i.e.}, $n=100$; see also
the violet curve in Fig.\,\ref{fig:poten_rejuven}). In this case,
the clusters remain finite but they become larger during cycles with
$\gamma_0=0.08$, shown in
Fig.\,\ref{fig:snapshots_rem5_n100}\,(a,\,c), while the deformation
is nearly reversible during subsequent relaxation at
$\gamma_0=0.06$, see Fig.\,\ref{fig:snapshots_rem5_n100}\,(b,\,d).
Hence, we conclude that the oscillatory deformation with alternating
excitation and relaxation periods leads to a controllable
rejuvenation without shear banding.

\vskip 0.05in

% nonaffine displacements; poorly annealed; continued

\begin{figure}[t]
\includegraphics[width=12.0cm,angle=0]{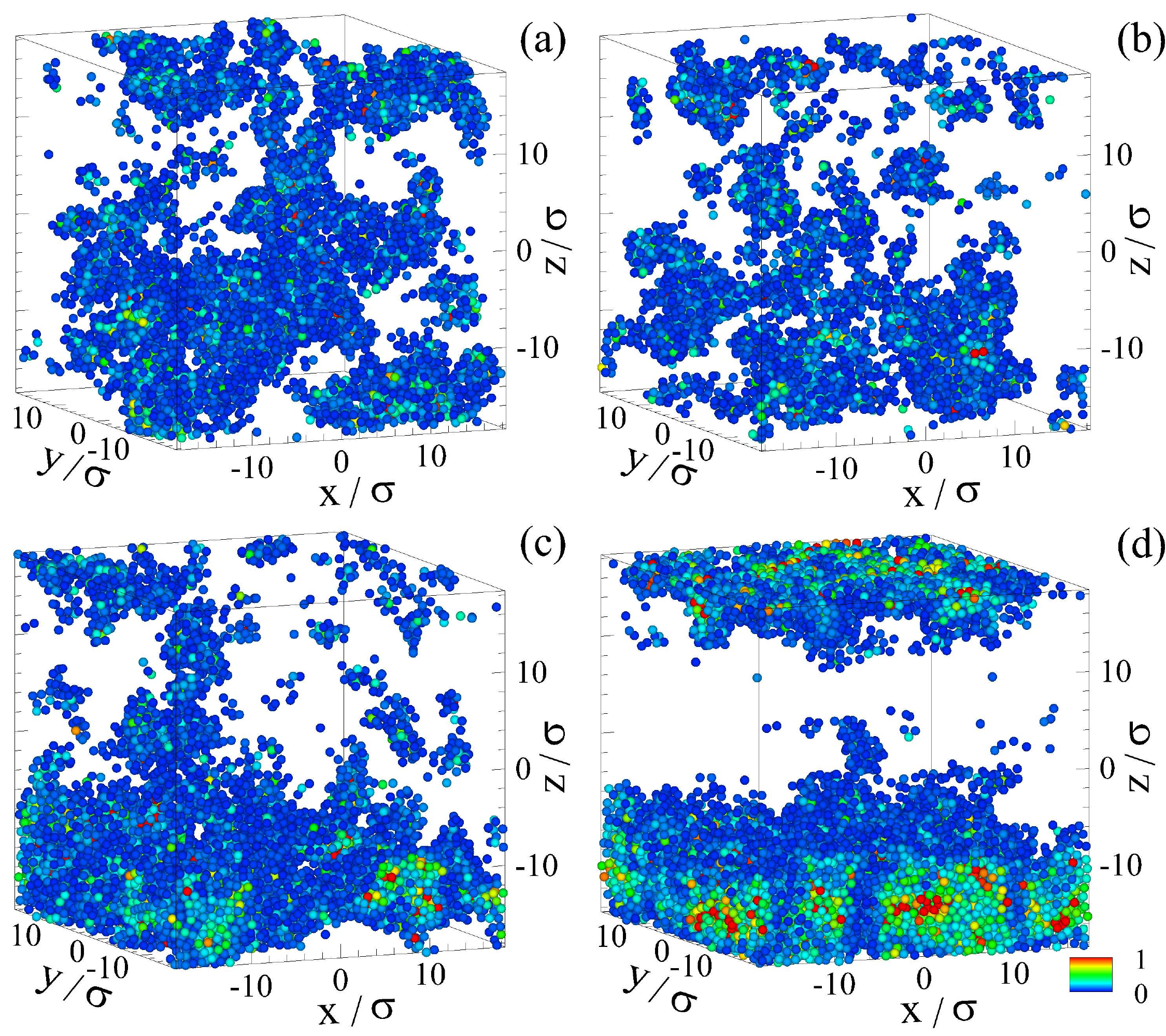}
\caption{(Color online) The sequence of snapshots of \textit{poorly
annealed} glass subjected to variable-amplitude periodic deformation
with periodicity $n=10$. The same loading protocol as in
Fig.\,\ref{fig:poten_relax} (the brown curve). The nonaffine measure
is (a) $D^2(100\,T, T)>0.04\,\sigma^2$, (b) $D^2(800\,T,
T)>0.04\,\sigma^2$, (c) $D^2(900\,T, T)>0.04\,\sigma^2$, and (d)
$D^2(1400\,T, T)>0.04\,\sigma^2$.  The colorcode in the legend
defines the magnitude of $D^2$.  }
\label{fig:snapshots_rem2_n10}
\end{figure}

\begin{figure}[t]
\includegraphics[width=12.0cm,angle=0]{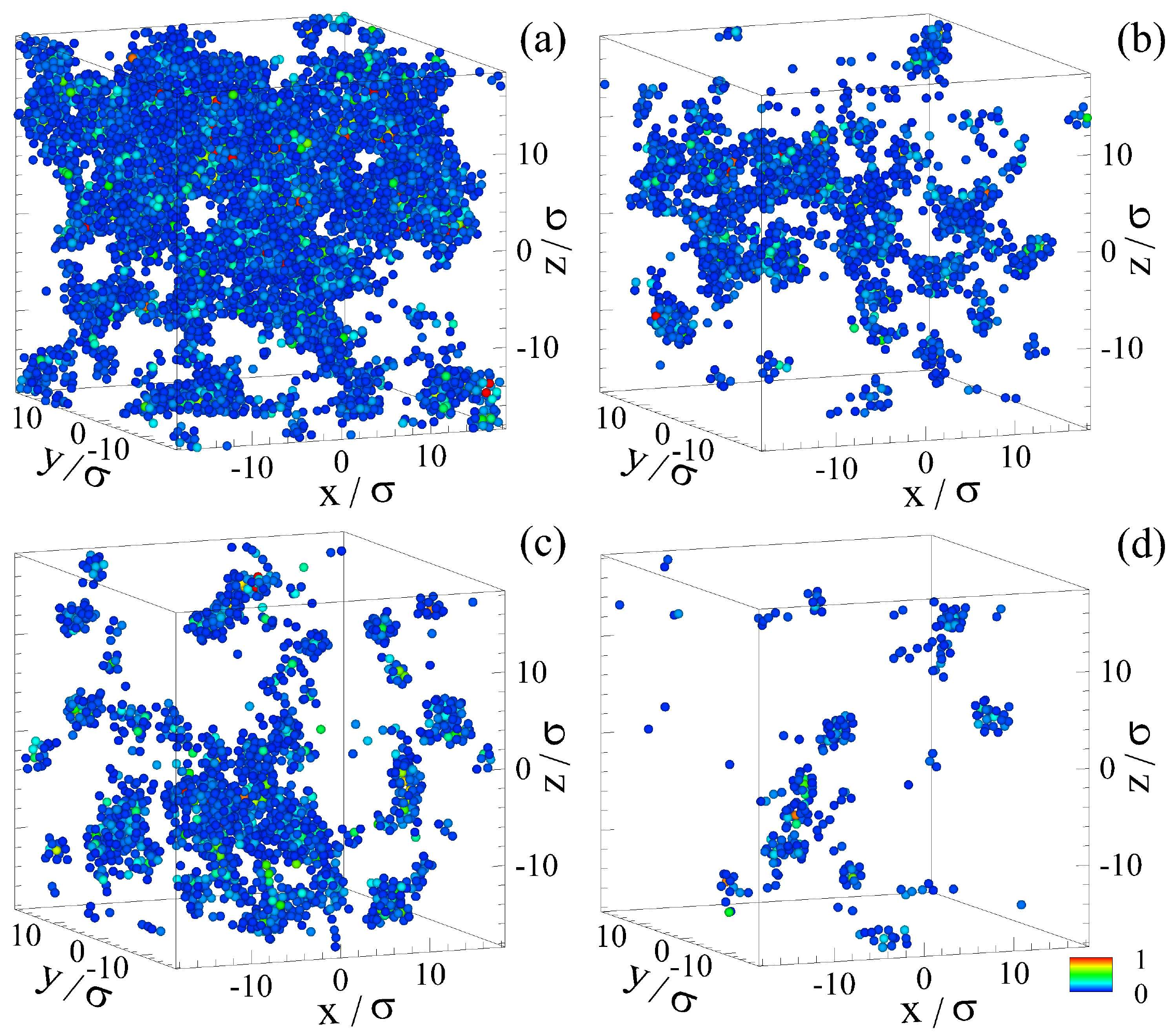}
\caption{(Color online) Spatial configurations of atoms with large
nonaffine displacements after one cycle for the \textit{poorly
annealed} glass under periodic deformation with $n=100$. See the red
curve in Fig.\,\ref{fig:poten_relax}.  The nonaffine quantity in
Eq.\,(\ref{Eq:D2min}) is (a) $D^2(100\,T, T)>0.04\,\sigma^2$, (b)
$D^2(150\,T, T)>0.04\,\sigma^2$, (c) $D^2(1900\,T,
T)>0.04\,\sigma^2$, and (d) $D^2(1950\,T, T)>0.04\,\sigma^2$. The
magnitude of $D^2$ is defined in the legend. }
\label{fig:snapshots_rem2_n100}
\end{figure}

Lastly, the sequences of stroboscopic snapshots of the
\textit{poorly annealed} glass under periodic deformation are
presented in Figures\,\ref{fig:snapshots_rem2_n10} and
\ref{fig:snapshots_rem2_n100} for $n=10$ and $100$, respectively.
The first two snapshots in
Fig.\,\ref{fig:snapshots_rem2_n10}\,(a,\,b) show interconnected
networks during 100 and 800-th cycles when the strain amplitude is
$\gamma_0=0.08$. These large-scale irreversible rearrangements
correspond to structural relaxation after rapid cooling, as
indicated by the brown curve in Fig.\,\ref{fig:poten_relax}. In this
loading protocol, the strain amplitude is changed to $\gamma_0=0.08$
frequently enough ($n=10$) to induce a yielding transition after
about 900 alternating cycles. Thus, one can clearly observe a single
shear band (across periodic boundaries) in
Fig.\,\ref{fig:snapshots_rem2_n10}\,(c,\,d) that is formed after
yielding and remains stable upon continued loading.  In the case
$n=100$, the \textit{poorly annealed} glass continues relaxation
during every 99 cycles with $\gamma_0=0.06$ [\,\textit{e.g.}, see
Fig.\,\ref{fig:snapshots_rem2_n100}\,(b,\,d)\,], which is
interrupted by one cycle with $\gamma_0=0.08$ that induces
large-scale plastic events, shown for example, in
Fig.\,\ref{fig:snapshots_rem2_n100}\,(a,\,c). The occasional
perturbation with the strain amplitude $\gamma_0=0.08$ facilitates
exploration of the potential energy landscape and relocates the
system to even lower energy states than cyclic loading with the
fixed amplitude $\gamma_0=0.06$ (see the red and black curves in
Fig.\,\ref{fig:poten_relax}).   Altogether, the results for well and
poorly annealed glasses under cyclic deformation with periodicity
$n=100$, shown in Figs.\,\ref{fig:snapshots_rem5_n100} and
\ref{fig:snapshots_rem2_n100}, demonstrate that glasses become
either highly rejuvenated or better annealed, thus extending the
range of accessible energy states.

\section{Conclusions}

In summary, molecular dynamics simulations were carried out to study
the effect of variable-amplitude periodic deformation on relaxation
and rejuvenation of disordered solids. The model glass former in
three dimensions was represented via the binary mixture, which was
cooled from the liquid state deep into the glass phase with
relatively slow and fast rates, producing well and poorly annealed
samples. The binary glass was subjected to cyclic loading where the
strain amplitude is fixed below the critical value but occasionally
changed slightly above the critical strain. During such deformation
protocol, one shear cycle with the large strain amplitude typically
induces collective plastic rearrangements but not shear bands if the
frequency of large-amplitude cycles is sufficiently low.

\vskip 0.05in

It was found that \textit{well annealed} samples can be rejuvenated
during a sequence of shear cycles provided that the strain amplitude
is rarely changed above the critical value. The increase in
potential energy is reflected in mechanical properties and leads to
enhanced ductility.  Interestingly, the same deformation protocol
drives \textit{poorly annealed} glasses to progressively lower
energy states, since large-amplitude cycles occasionally perturb the
system and prevent trapping in local minima of the potential energy
landscape. On the other hand, both well and poorly annealed glasses
undergo a yielding transition after a number of transient cycles
when the strain amplitude is frequently changed above and below the
critical value. These conclusions were confirmed by visualizing
spatial configurations of atoms with large nonaffine displacements
that are organized either in compact clusters or planar shear bands.

\section*{Acknowledgments}

Financial support from the National Science Foundation (CNS-1531923)
is gratefully acknowledged.  The article was prepared within the
framework of the HSE University Basic Research Program and funded in
part by the Russian Academic Excellence Project `5-100'. The
simulations were performed at Wright State University's Computing
Facility and the Ohio Supercomputer Center using the LAMMPS code
developed at Sandia National Laboratories~\cite{Lammps}.

% \section*{Conflict of Interest}
% The author declares that he has no conflict of interest.

%%%%%%%%%%%%%%% FIGURES %%%%%%%%%%%%%%%%%%%%%%%

% definition of the modulated amplitude
%

% potential energy minima rejuvenation yielding
%

% potential energy minima relaxation yielding
%

% G and Y versus #cyc for rejuvenated glass n=100
%

% snapshots xz for well annealed glass n=20
%

% snapshots xz for well annealed glass n=100
%

% snapshots xz for poorly annealed glass n=10
%

% snapshots xz for poorly annealed glass n=100
%

\bibliographystyle{prsty}

\end{document}